\documentstyle[12pt,fleqn,epsf]{article}
\renewcommand{\baselinestretch}{1.2}

\def\fnote#1#2{\begingroup\def\thefootnote{#1}\footnote{#2}\endgroup}
\makeatletter
\def\section{\@startsection {section}{1}{\z@}{3.5ex plus 1ex minus
    .2ex}{2.3ex plus .2ex}{\sc }}
\def\subsection{\@startsection{subsection}{2}{\z@}{3.25ex plus 1ex
minus
   .2ex}{1.5ex plus .2ex}{\small \sc }}
\def\appendix{\par\clearpage
  \setcounter{section}{0}
  \setcounter{subsection}{0}
  \@addtoreset{equation}{section}
  \def\@sectname{Appendix~}
  \def\theequation{\thesection.\arabic{equation}}
  \def\thesection{\Alph{section}}}
\makeatother
\def\ap#1#2#3{     {\it Ann. Phys. (NY) }{\bf #1} (19#2) #3}

\def\npb#1#2#3{    {\it Nucl. Phys. }{\bf B #1} (19#2) #3}
\def\plb#1#2#3{    {\it Phys. Lett. }{\bf B #1} (19#2) #3}
\def\prd#1#2#3{    {\it Phys. Rev. }{\bf D #1} (19#2) #3}

\def\zpc#1#2#3{    {\it Z. Physik }{\bf C #1} (19#2) #3}

\def\eq#1{{eq.~(\ref{#1})}}
\def\eqs#1#2{{eqs.~(\ref{#1})--(\ref{#2})}}

\let\vev\VEV

\def\Tr{\mathop{\mbox{Tr}}\,}
\def\etal{{\it et al.}}

\newcommand{\bea}{\begin{eqnarray}}
\newcommand{\beq}{\begin{equation}}
\newcommand{\eea}{\end{eqnarray}}
\newcommand{\eeq}{\end{equation}}
\newcommand{\nnu}{\nonumber}
\newcommand{\spav}[1]{\parbox{1mm}{\vspace*{#1}}}

\begin{document}
\begin{titlepage}
\begin{flushright}
{\tt SISSA 74/96/EP}
\end{flushright}
\spav{1cm}
\begin{center}
{\Large\bf The Bosonization of the}\\
{\Large\bf Electroweak Penguin Operators}\\
\spav{1cm}\\
{\large M. Fabbrichesi and
E.I. Lashin\fnote{\dag}{Permanent address:
Ain Shams University, Faculty of Science, Dept. of Physics, Cairo, Egypt.}}
\spav{1.5cm}\\
{\em  INFN, Sezione di Trieste}\\
{\em and}\\ 
{\em Scuola Internazionale Superiore di Studi Avanzati}\\
{\em via Beirut 4, I-34013 Trieste, Italy.}\\
\spav{1.5cm}\\
{\sc Abstract}
\end{center}
We give the complete $O(p^2)$ bosonization of the electroweak Penguin
operators $Q_{7,8}$ and compare the result
with that of the gluon Penguin operators $Q_{5,6}$. We find that, 
in addition to the
usual (constant and current-current) parts, 
there are three new terms not discussed previously in the literature.
Two of these are present  in the factorization approximation and should 
be included in the standard definition of the $B_{7,8}$-factors.  
The impact of these corrections
on the direct $CP$-violating parameter $\varepsilon '/\varepsilon$ 
is briefly discussed.
\vfill
\spav{.5cm}\\
{\tt SISSA 74/96/EP}\\
{\tt  May 1996}

\end{titlepage}

\newpage
\setcounter{footnote}{0}
\setcounter{page}{1}

{\bf 1.} Penguin operators play an important role in the physics of
the standard model at low energies and, in particular, in the non-leptonic
decays of kaons. The estimate of their matrix elements has been a controversial
subject from the very beginning because of a subtle cancellation of the
leading contribution in chiral perturbation theory.

Most of the debate~\cite{debate,1/N}, however, was centered around the 
gluon Penguin operators---at the
time the only relevant ones---that we write as
\bea
Q_{5} & = & \left( \overline{s} d \right)_{\rm V-A}
   \sum_{q} \left( \overline{q} q \right)_{\rm V + A}  \\
Q_{6}  & = & \left( \overline{s}_{\alpha} d_{\beta}  \right)_{\rm V-A}
   \sum_{q} ( \overline{q}_{\beta}  q_{\alpha} )_{\rm V+A}
\eea
by means of the by-now-standard notation in which
$\alpha$, $\beta$ denote color indices ($\alpha,\beta
=1,\ldots,N_c$), the subscripts $(V\pm A)$ refer to the chiral projections
$\gamma_{\mu} (1 \pm \gamma_5)$ and color
indices for the color singlet operators are omitted. The other gluon Penguin operators ($Q_{3,4}$) 
discussed in
the literature are readily bosonized as the product of two left-handed
currents and we will not discuss them.
 
The electroweak Penguin 
operators~\cite{Lisa} are defined as
\bea
Q_{7} & = & \frac{3}{2} \left( \overline{s} d \right)_{\rm V-A}
         \sum_{q} \hat{e}_q \left( \overline{q} q \right)_{\rm V +A} \\
Q_{8} & = & \frac{3}{2} \left( \overline{s}_{\alpha}
          d_{\beta} \right)_{\rm V-A}
     \sum_{q} \hat{e}_q ( \overline{q}_{\beta}  q_{\alpha})_{\rm V+ A}\, ,
\eea
where
$\hat{e}_q$  are the quark charges ($\hat{e}_d =
\hat{e}_s = - 1/3$ and $\hat{e}_u = 2/3$). The appereance of these
operators has not revived the discussion on their bosonization
and they
 have been treated along the same lines as the gluon ones.
Yet, new features emerge when considering this new class 
and the argument followed in the bosonization and the
determination of the chiral coefficient of the gluon Penguin fail us here. 
Let us see why.

\bigskip \bigskip
{\bf 2.} In order to proceed we first write the two electroweak operators as
\bea
Q_7 & = & \frac{3}{2} \hat{e}_d Q_5 + \frac{3}{2} ( \hat{e}_u -
\hat{e}_d ) \Delta Q_7 \nnu \\
Q_8 & = & \frac{3}{2} \hat{e}_d Q_6 + \frac{3}{2} ( \hat{e}_u -
\hat{e}_d ) \Delta Q_8 
\eea
where
\bea
\Delta Q_7 & = & \left( \overline{s} d  \right)_{\rm V-A}
            \left( \overline{u}  u \right)_{\rm V+A} \nnu \\
\Delta Q_8 & = & \left( \overline{s}_{\alpha} d_{\beta}  \right)_{\rm V-A}
            \left( \overline{u}_{\beta}  u_{\alpha} \right)_{\rm V+A} \, .
\eea
This splitting has the advantage of separating out the pure octet part of the
electroweak operators---
that is,  the  part already present in the case of the gluon
Penguins and which therefore does not introduce any new features.

To obtain the chiral representation of the operator $\Delta Q_{7,8}$ we must consider
the bosonization to $O(p^2)$ of the quark densities:
\bea
 \bar{q}^j_L q^i_R   & \to &  - 2 B_0
\left[ \frac{f^2}{4} \Sigma + L_5 \ \Sigma
D_\mu \Sigma^{\dag} D^\mu \Sigma +
4 B_0 L_8 \ \Sigma {\cal M}^\dag\ \Sigma \right]_{ij} \nnu \\
 \bar{q}^j_R q^i_L   & \to &  - 2 B_0
\left[ \frac{f^2}{4} \Sigma^\dag + L_5 \ \Sigma^\dag
D_\mu \Sigma D^\mu \Sigma^{\dag} +
4 B_0 L_8 \Sigma^{\dag} {\cal M} \ \Sigma^{\dag} \right]_{ij} \, , 
\label{qLqR}
\eea
where $B_0 = - \vev{\bar q q}/f^2$, ${\cal M} = \mbox{diag}[m_u,m_d,m_s]$
and the subscript $ij$ is the flavor projection. The bosonization in \eq{qLqR} can be obtained
by either considering directly the generalized mass term in the strong lagrangian
or by computing
\beq
 \frac{\delta {\cal L}^{\chi PT}}{\delta {\cal M}^\dag} \qquad \mbox{and}
\qquad  \frac{\delta {\cal L}^{\chi PT}}{\delta {\cal M}}
\eeq
respectively, where ${\cal L}^{\chi PT}$ is the strong lagrangian to $O(p^4)$.  

The possible terms containing
 second derivatives have been 
eliminated in \eq{qLqR} by means of the equations of motion. 
If we do not use the equations of motion and retain the
second derivative terms there are two additional terms that in principle must
be added 
\beq
2  B_0 \,c_1 \: D^2 \Sigma  \: + \: 2  B_0 \,c_2 \:
 \Sigma D^2 \Sigma \Sigma^\dag \, . \label{dd}
\eeq
However, only two out of the three derivative terms  in \eq{qLqR} and \eq{dd}
are independent
because we can eliminate
one of them by means of the relation
\beq
D^2 \Sigma + \Sigma D^2 \Sigma \Sigma^\dag + 2 D_\mu \Sigma
D^\mu \Sigma^\dag \Sigma = 0 \, ,
\eeq
which follows by the unitarity of the matrix $\Sigma$ alone. 
The addition of these terms does not
modify our result based on \eq{qLqR} except for a non-factorizable
term~\cite{HME} 
\beq
c_{\rm nf} \Tr \left( \lambda^3_2 D_\mu  \Sigma \: \lambda^1_1 
 D^\mu \Sigma^{\dag} \right)  \label{nf} 
\eeq
which is generated by the second order derivative acting on the
two densities simultaneously. Such a term goes beyond  the vacuum
saturation approximation (VSA)  
and we shall neglect it.

Additional terms proportional to the counterterms $L_4$, $L_6$ and $L_7$ 
are not necessary because they  give  contributions to \eq{qLqR}  
of the form, for instance for $L_4$:
\beq
 2  B_0 \, L_4 \: \Tr (D^\mu \Sigma^\dag D_\mu \Sigma ) \Sigma_{ij} \, ,
\eeq
which only represent a
 wave-function renormalization
induced by the strong sector.

We prefer to use the equations of motion because we only need the higher order
terms at the tree (on-shell) level and---as stressed 
in~\cite{Farese}---the counterterms necessary in the
renormalization procedure are only correct if a classical background is assumed.
 This is the same procedure
followed in the definition of the $O(p^4)$ chiral lagrangian 
${\cal L}^{\chi PT}$ on which
\eq{qLqR} is based. 

The crucial 
point is that---no matter what procedure one follows---there are 
still two independent
terms in \eq{qLqR} beside the leading (constant) one.

We now turn to the bosonization of the electroweak operator $Q_8$.
By applying \eq{qLqR} to the  operator
\beq
 - 12 (\overline{s}_L u_R )(\overline{u}_R d_L)
\eeq
which is  obtained by a Fierz transformation from
the operator $3 \Delta Q_8/2$,
we obtain three terms
\bea
 & &
- 3 \left( \vev{\bar q q} \right)^2
\Tr \left( \lambda^3_1 \Sigma^{\dag} \right) \Tr \left( \lambda^1_2 \Sigma
\right) \label{14} \\
& &  - \, 12  \frac{\vev{\bar q q}^2 L_5}{f^2} \left[ 
\Tr \left( \lambda^3_1 \Sigma \right)
\Tr \left( \lambda^1_2  D_\mu
\Sigma^{\dag} D^\mu \Sigma\ \Sigma^{\dag} \right) \right.  \nnu \\
& &  \quad \quad \quad \quad \quad + \left.
\Tr \left( \lambda^3_1  D_\mu \Sigma D^\mu \Sigma^{\dag}\ \Sigma\right)
\Tr \left(  \lambda^1_2 \Sigma^{\dag} \right)  \right] \\
& & - \,48 \frac{\vev{\bar q q}^2 B_0 L_8}{f^2} \ 
\left[ \Tr \left( \lambda^1_2 \Sigma^\dag\right) \Tr \left(\lambda _1^3 
\Sigma {\cal M}^\dag \Sigma \right)\right.  \nnu \\
& &  \quad \quad \quad \quad \quad + \left.
\Tr \left(\lambda_1^3 \Sigma \right) \Tr \left( \lambda^1_2  \Sigma^\dag 
{\cal M} \Sigma^\dag \right)\right] \, ,  \label{16}
\eea
where the projection matrices are defined by
 $(\lambda^i_j)_{lk} = \delta_{il}\delta_{jk}$. 
Another term
\beq
 -\frac{3 f_\pi^4}{2} \: \Tr \left( \lambda^3_2 \Sigma^{\dag} D_\mu
\Sigma \right)
\Tr \left(  \lambda^1_1 \Sigma D^\mu \Sigma^{\dag} \right)  \label{17} 
\eeq
is obtained by considering the leading order 
bosonization of the quark currents in the operator $Q_8$ as
\bea
q^j_L\gamma^\mu q^i_L & \rightarrow & -i \frac{f_\pi^2}{2} \left( \Sigma^{\dag} D_\mu
\Sigma \right)_{ij} \\
q^j_R\gamma^\mu q^i_R & \rightarrow & -i \frac{f_\pi^2}{2}  \left( \Sigma D_\mu
\Sigma^{\dag} \right)_{ij} \, .
\eea
Such a term is completely determined, being of the leading order, and requires
no further discussion.

The current literature on the VSA estimate of the electroweak
operators~\cite{Lisa} has dealt only with the terms (\ref{14}) and 
(\ref{17}). Ref.~\cite{HME} discusses the complete lagrangian by means of
a different bosonization technique which also include non-factorized
configurations. In comparing our lagrangians with that written in~\cite{HME}, 
care should be taken in rewriting single traces as double
traces and vice versa (see the appendix of ref.~\cite{HME}). 

The usual expression for the bosonization of the pure octet
 part (the only part for the gluon Penguins $Q_{5,6}$) is readily
 obtained from \eqs{14}{16} by replacing the projection on
 the $u$ (1) quark by the 
 sum over all quark flavors to obtain:
 \bea
& & -24 \:\frac{\vev{\bar q q}^2 L_5}{f^2}
\Tr \left( \lambda^3_2 D_\mu \Sigma^{\dag}
D^\mu \Sigma
\right)  \label{19} \\
& & - 48 \: \frac{\vev{\bar q q}^2 B_0 L_8}{f^2} \ 
 \Tr \left[  \lambda^3_2  \left( {\cal M}^\dag \Sigma + 
 \Sigma^\dag {\cal M} \right) \right]  \, , 
 \eea 
where the leading constant term has vanished by projection
 and the mass correction is just
 a renormalization that can  be reabsorbed in the definition of
$\chi = 2 B_0 {\cal M}$ in the strong lagrangian. 

In this case there is only one possible bosonization term (\ref{19}) and
one constant ($L_5$) to 
be determined. The usual bosonization for $Q_6$ (and \eq{Q6} below)
 is obtained by multiplying
\eq{19} by 2/3.

\bigskip \bigskip
{\bf 3.} If we neglect the non-factorized term(\ref{nf}) and 
the mass term (\ref{16}, then the 
chiral lagrangian necessary in the VSA approximation is determined by the
same argument used in the case of the gluon Penguin. 
In particular, it is possible to fix $L_5$
by computing the ratio between the kaon and pion decay constants:
\beq
\frac{{\cal A} ( K^+ \rightarrow \mu^+ \nu_\mu)}
{{\cal A} ( \pi^+ \rightarrow \mu^+ \nu_\mu)} \, 
\eeq
which gives, in the large $1/N_c$ limit~\cite{1/N},
\beq
L_5 = \frac{1}{4}
\left( 1 - \frac{f_K}{f_\pi} \right) \frac{f_\pi^2}{m_K^2 - m_\pi^2}
\eeq
and
\beq
L_5 (m_\rho) = (1.4 \pm 0.5) \: 10^{-3} \, ,
\eeq
if chiral logarithms are kept~\cite{GL}.

In order to include the mass-term correction,
the constant $L_8$ can be determined from the
GMO formula and the knowledge of $L_5$. It is found~\cite{GL} that
\beq
L_8 (m_\rho) = (0.9 \pm 0.3) \: 10^{-3} \, .
\eeq

We now write the matrix elements of the operator $Q_6$ and $Q_8$ (those of
$Q_5$ and $Q_7$ are readily obtained by going to the next order in $1/N_c$)
for the decay of a kaon into two pions.
As usual, we split them into isospin amplitudes ($I= 0,2$) and find
\beq
\vev{Q_6}_0^{\rm VSA} = - 4 \frac{\vev{\bar q q}^2}{f^4 \Lambda^2} X \label{Q6}
\eeq
where $X =\sqrt{3} f_\pi ( m_K^2 - m_\pi^2)$.  Eq.(\ref{Q6}) is the usual VSA
result for the gluon Penguin operator.

The operator $Q_8$ dominates the channel $I=2$ where we find that
\beq
\langle Q_8 \rangle _2^{\rm VSA} = 
\sqrt{6} \, \frac{\langle \bar{q} q \rangle ^2}{f_\pi^3}
  + 4 \sqrt{6} \, \frac{\vev{\bar{q}q} ^2}{f_\pi^5}  
  \left( 4 L_8  - L_5 \right) m_K^2 
   - \frac{\sqrt{2}}{2 N_c} X \label{Q8}
   \eeq
 The second term in \eq{Q8}  is the result of the two new terms we are
 discussing. It is as
 large as 20\% of the leading one, as opposed to the small momentum correction
 contained in the third and last term that is only 1\%.
 
 As we have already pointed out,
 in the literature~\cite{Lisa}, only the leading (constant) term and the $X$-term
 have been so far included. As a consequence, the usual definition of the 
 $B_8$-factor
 \beq
 B_8 \equiv \frac{\vev{Q_8}^{\rm any \: model}}{\vev{Q_8}^{\rm VSA}} \, ,
 \eeq
  that quantifies deviation from the VSA is not correct and  should be changed
 according to \eq{Q8}.
 
\bigskip \bigskip
{\bf 5.} Because of the importance of electroweak operators in the
determination of the direct $CP$-violating parameter $\varepsilon'/\varepsilon$,
we have estimated their effect in the VSA and compared the result
that includes the new terms with that that does not.
As it can be seen in Fig. 1, the effect is about 20\% in the range of 
input parameters we considered.  
\begin{figure}[htb]
\epsfxsize=12cm
\centerline{\epsfbox{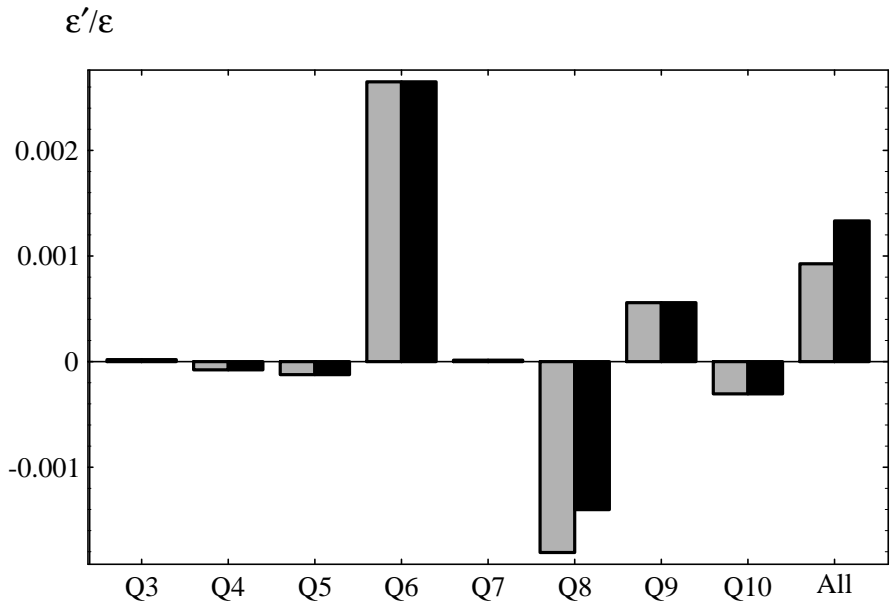}}
\caption{Contributions of the various operators in the VSA to
$\varepsilon'/\varepsilon$ at $\mu =0.8$ GeV. Grey (black) bars (do not) include
the extra terms. All operators are shown for completeness.}
\end{figure}
We should
however bear in mind that the cancellation between gluon and electroweak 
operators
is not as effective
in the VSA as it is in all  more refined computations~\cite{eps,BEF}, as
shown by the rather large final value of $\varepsilon'/\varepsilon$ in Fig.1.
For this reason, in any computation in which such a cancellation is more
complete,
 the impact of the new terms could be much more dramatic

\bigskip \bigskip
{\bf 6.} For the sake of comparison, we now turn to a specific example
of modeling of low-energy QCD. In the chiral quark model
($\chi$QM) (see ref.~\cite{HME} for discussion
and a list
of references) we have 
\bea
L_5^{\chi QM} & = &  \frac{ f^4}{8 M |\vev{\bar q q}|}
\left( 1 - 6 \frac{M^2}{\Lambda_{\chi}^2} \right) \\
L_8^{\chi QM} & = &   \frac{ f^4}{16 M |\vev{\bar q q}|} \left( 1 - 
\frac{M f^2}{|\vev{\bar q q}|} \right) - \frac{1}{24} \,
\frac{N_c}{16 \pi^2}
\eea
where $M$ is a parameter characteristic of the model.

We thus find that
\beq
\langle Q_8 \rangle _2 ^{\rm \chi QM}= 
\sqrt{6} \, \frac{\langle \bar{q} q \rangle ^2}{f_\pi^3}
  + 4 \sqrt{6} \, \frac{\vev{\bar{q}q} ^2}{f_\pi^5} 
  \left[ c_{\rm nf}\, m_\pi^2 + \left( 4 L_8^{\rm \chi QM} - 
  L_5^{\rm \chi QM} \right)\, m_K^2 \right]
  - \frac{\sqrt{2}}{2 N_c} X 
\label{Q8Y} \, ,
\eeq
where also the non-factorizable term proportional to
\beq 
c_{\rm nf} = \frac{1}{4} \, \frac{f^4}{M|\vev{\bar q q}|}
\eeq
 is included for completeness.

In particular, the combination 
\beq
4 L_8^{\chi QM} - 
  L_5^{\chi QM} = \frac{3}{8}\,\frac{f^4}{M |\vev{qq}|} 
  -\frac{3}{4}\,\frac{ f^4 M}{\Lambda^2 |\vev{\bar q q}|}  
  -\frac{1}{4}\,\frac{f^6}{|\vev{qq}|^2}  - \frac{1}{8} \,\frac{N_c}{16 \pi^2}
  \label{q8QM}
\eeq 
  turns out to be numerically of the same order as  the VSA result. An  estimate
  of $\varepsilon'/\varepsilon$ in this model (albeit in the
  chiral limit inclusive of only the 
  second of the three terms in \eq{q8QM} and of the one proportional to
   $c_{\rm nf}$) is presented in ref.~\cite{BEF}. 
  
\bigskip \bigskip
M.F. would like to thank E. de Rafael for discussions.

%
%
\clearpage
\renewcommand{\baselinestretch}{1}


\begin{thebibliography}{99}
{\small

\bibitem{debate} M.A. Shifman, A.I. Vainsthain and V.I. Zakharov,
\npb{120}{77}{316};\\
J.F. Donoghue \etal, \prd{21}{80}{186};\\
J. F. Donoghue, \prd{30}{84}{1499};\\
A.J. Buras and J.-M. G\'{e}rard, \npb{264}{86}{371};\\
W.A. Bardeen A.J. Buras and J.-M. G\'{e}rard, \plb{192}{87} 138, 156;\\
G. Buchalla, A.J. Buras and K. Harlander, \npb{337}{90}{313}.

\bibitem{1/N} R.S. Chivukula, J.M. Flynn and H. Georgi, \plb{171}{86}453;

\bibitem{GL}J. Gasser and H. Leutwyler, \ap{158}{84}{142},
\npb{250}{85}{465,517,539}. 

\bibitem{Lisa} J. Bijnens and M.B. Wise, \plb{137}{84}{245};\\
J. Flynn and L. Randall, \plb{224}{89}{221}; Erratum, \plb{235}{90}{412};\\
M. Lusignoli, \npb{325}{89}33;\\
G. Buchalla, A.J. Buras and M.K. Harlander, \npb{337}{90}{313}.

\bibitem{HME} V. Antonelli, S. Bertolini, J.O. Eeg,
M. Fabbrichesi and E.I. Lashin, {\it The $\Delta S = 1$ Weak Chiral
Lagrangian as the Effective Theory of the Chiral Quark Model}, preprint
SISSA 43/95/EP (September 1995), hep-ph/9511255, to appear in 
{\it Nuclear Physics} {\bf B}.

\bibitem{Farese} G. Esposito-Farese, \zpc{50}{91}{255}.

\bibitem{eps} M. Ciuchini, E. Franco, G. Martinelli and L. Reina,
{\it Estimates of $\varepsilon'/\varepsilon$}, in {\it The Second
DA$\Phi$NE Physics Handbook}, eds. L. Maiani et al. (Frascati, 1995);
\zpc{68}{95}{239} and references therein; \\
 G. Buchalla, A.J. Buras and M.E. Lautenbacher, 
{\em Weak Decays beyond Leading Logarithms}, hep-ph/95112380, to appear
in {\em Rev. Mod. Phys.} and references therein.

\bibitem{BEF}  S. Bertolini, J.O. Eeg and
M. Fabbrichesi, {\it A New Estimate of $\varepsilon'/\varepsilon$}, preprint
SISSA 103/95/EP (November 1995), hep-ph/9512356, to appear in 
{\it Nuclear Physics} {\bf B}.



}
\end{thebibliography}
\end{document}